\input harvmac
\input epsf
\vskip 1cm

 \Title{ \vbox{\baselineskip12pt\hbox{  Brown Het-1226 }}}
 {\vbox{
\centerline{ A remark on T-duality  }
\centerline{ and quantum volumes of zero-brane moduli spaces.  }  }}

\centerline{$\quad$ { Sanjaye Ramgoolam }}
\smallskip
\centerline{{\sl  }}
\centerline{{\sl Brown  University}}
\centerline{{\sl Providence, RI 02912 }}
\centerline{{\tt ramgosk@het.brown.edu}}
\vskip .3in 

 T-duality ( Fourier-Mukai duality ) and properties 
 of classical instanton moduli spaces can be used to 
 deduce some properties of $\alpha^{\prime}$-corrected 
 moduli spaces of branes for Type IIA string theory 
 compactified on  $K3$ or $T^4$.  Some interesting
 differences between the two compactifications 
 are exhibited.

\lref\zam{ Zamolodchikov } 
\lref\ahberk{ Aharony and Berkooz }  
\lref\grgu{ M. Green and M. Gutperle } 
\lref\yosh{ K. Yoshioka, ``Irreducibility of moduli spaces of vector bundles
 on $K3$ surfaces, math.AG/9907001. }
\lref\witade{ E. Witten, ``Heterotic string on ADE'' } 
\lref\bss{ Banks, Seiberg and Silverstein } 
\lref\sw{ N. Seiberg and E. Witten, ``Non-commutative geometry''} 
\lref\mms{ Moore, Marino, Witten} 
\lref\df{ Danielsson, Ferretti } 
\lref\dougsm{ M. Douglas, ``Superstring dualities, Dirichlet branes,
 and the small scale structure of space-time,'' hep-th/9610041 } 
\lref\shenl{ S. Shenker }
 \lref\ramwal{ S. Ramgoolam and D. Waldram, ``Zero branes on a compact 
orbifold,'' JHEP-9807; 009,1998;  hepth/9805191. } 
\lref\nwen{ W. Nahm and K. Wendland, ``A hiker's guide to K3: Aspects 
of $N=(4,4)$ CFT with central charge $c=6$ ,'' hep-th/9912067. } 
\lref\asp{ P. Aspinwall, ``K3 surfaces and string duality,'' hep-th/9611137 } 
\lref\mikh{ A. Mikhailov }
\lref\dij{ R. Dijkgraaf }  
\lref\hamo{ J. Harvey and G. Moore, ``On the algebras of BPS states,''
   Commun.Math.Phys.197:489-519,1998;   hep-th/9609017. }
 \lref\mr{ M. Mihailescu and S. Ramgoolam, 
``Duality and the combinatorics of long strings in ADS3,'' hep-th/0002002 } 
\lref\grka{ B. R. Greene and Y. Kanter, 
``Small volumes in Compactified String theory,''hep-th/9612181, 
 NPB497 (1997) 127-145. }
\lref\doog{ M. Douglas and H. Ooguri, ``Why Matrix theory is hard,'' 
hep-th/9710178, Phys. Lett. B425(1998) 71-76. }
\lref\kop{  E. Kiritsis, N. Obers and  B. Pioline,  
``Heterotic/Type II Triality and Instantons on $K_3$,'' hep-th/0001083, 
 JHEP 0001 (2000) 029 } 
\lref\bre{ I. Brunner, R. Entin and C. Romelsberger, `` D-branes on
$T^4/Z_2$ and T-duality,'' JHEP 9906 (1999) 16,  hep-th/9905078 }

\Date{June 2000}

\newsec{ Introduction } 

 The properties of space-time in string theory 
 are very mysterious, since space-time often arises 
 as a derived object from more primitive concepts, 
 e.g worldsheet string theory, or the moduli space 
 of vacua of a gauge theory. As such it manifests 
 properties quite different
 from expectations based on ordinary classical geometry.  
 Such properties include T-duality where 
 a theory defined on a large circle has the same 
 physics as a theory defined on a small circle, 
 and space-time non-commutativity where the theory 
 shows evidence of non-commuting coordinates. 
 In this note, we explore some aspects of large-small 
 dualities and observe their consequences for moduli spaces of 
  zero branes on $K3$ and $T^4$. We work with units where 
 $4 \pi^2 \alpha^{\prime} = 1 $, so that T-duality takes
 $V \rightarrow 1/V$.   

\newsec{ Fourier-Mukai and  
Quantum volumes 
          of brane moduli spaces } 

      There is a T-duality symmetry in the 
      $O(4,4;Z)$ T-duality group of Type IIA on $T4$ 
      which inverts the volume of the 4-torus, 
      in the absence of B-fields. There is also 
      such a symmetry in $O(4,20;Z)$, the duality group 
      of a $K3$. Let us recall the set-up which is used to describe such  
   a duality. 

 The moduli space of  positive 4-planes  in a Lorentzian space
 $H^*(K3,R) = R^{(4,20)}$
 describes the moduli space of compactifications 
 of type IIA on $K3$ \asp. Let us label basis vectors 
 spanning the 4-plane as $E^{1}$ to $E^{4}$. They can be expressed
 in terms of the moduli as     
\eqn\modvec{\eqalign{& E^{1} = (V - {1 \over 2 }  B.B , 1; 0 )  \cr 
                     & E^{i} = ( 0, - B.\omega^i; \omega^{i} ) \cr }}
 The vectors $\omega^{i}$ are self-dual 2-forms living in the 
 lattice $H^2(K3,R) = R^{3,19}$, 
 and describe the moduli space of Einstein metrics of 
 unit volume. 
   The space of inequivalent compactifications 
 is a discrete quotient of the Grassmannian $O(4,20)/O(4) \times O(20)$ : 
\eqn\quot{ O(4,20;Z)\backslash O(4,20)/O(4) \times O(20),  } 
since physical quantities depend on a choice of 
$(0,2, {\hbox {or} }~  4)$-brane 
charge in the lattice $\Gamma^{(4,20)} \subset R^{(4,20)} $ 
and a choice of background. 
The discrete quotient is by symmetries of the lattice. 
 In the absence of B-fields we can invert the volume of
 the $K3$ by a transformation which involves permuting the first 
 two entries of the vectors \asp. 
The usual T-duality inverting the
 volume of the torus can also be described in a 
 similar language with
 $\Gamma^{(4,4)} \subset R^{(4,4)} $
 replacing $\Gamma^{(4,20)} \subset R^{(4,20)}$.
 These dualities are called Fourier-Mukai dualities
 in view of their action  on the gauge theory 
 describing the dual brane systems.  
 A different element of the T-duality 
 group  inverts the volume of $K3$ in the presence of 
 special B-fields present in the perturbative orbifold limit 
 of $K3$ \ramwal\nwen\kop\bre. The case $B=0$ will be of interest here.

\subsec{ Zero brane on $T^4$ }
      We will consider the consequences of this 
  duality on a system of zero-brane 
      on $T^4 $ with $B=0$.  The moduli space of a zero-brane on $T^4 $
     in the large volume limit can safely be said to be identical
    to the $T^4$ itself. All the K\"ahler and complex structure 
      parameters of the moduli space of the zero-brane are identical 
      to those of the base space itself.

 Let us denote by $M_{(Q_4,Q_0)}(X)$ the moduli space 
 of $Q_0$ zero-branes and $Q_4$ 4-branes on $X$, 
 where $X$ is $T4$, a four-torus, or $K3$.
  The system $(Q_4,Q_0)$ is associated with the Mukai 
 vector $(Q_4, Q_0-Q_4)$ in the case of $K3$ ,
 and $(Q_4,Q_0)$ in the case of $T^4$ \hamo. 
   
   The quantity of immediate  interest will be 
   the volume of the moduli space of 
 a zero-brane on $T^4 $, which we will 
 denote as $Vol ( M_{0,1} ( T (V) ) )$. 
 In the large volume limit, by the above reasoning 
 $Vol ( M_{0,1} ( T (V) ) ) = V$.  
   
 Now consider varying the geometry of the torus, 
 reducing its volume while keeping all other
 K\"ahler and complex structure parameters fixed. 
 Once we reach the small volume region, we can use 
 T-duality to map to the large volume region, 
 and at the same time map the 0-brane to 4-brane.
 By considering the moduli space of the 4-brane
 in the large volume limit we can learn about the 
 moduli space of the zero-brane in the small volume limit. 

 The 4-brane in the large volume limit is 
  described by $U(1)$ gauge theory.
 The moduli space is  the space of flat connections
 on the torus. Since the moduli space of flat connections 
 on the large torus is the dual torus which has small volume,
 we deduce that the moduli space of the zero brane as a function of $V$ 
 behaves as  
 $ Vol ( M_{1,0} ( T ( 1/V ) ) ) = V $ in the region of small $V$,  

\fig\volmodt{ 
{Fig.1 }  } 
{\epsfxsize3.0in\epsfbox{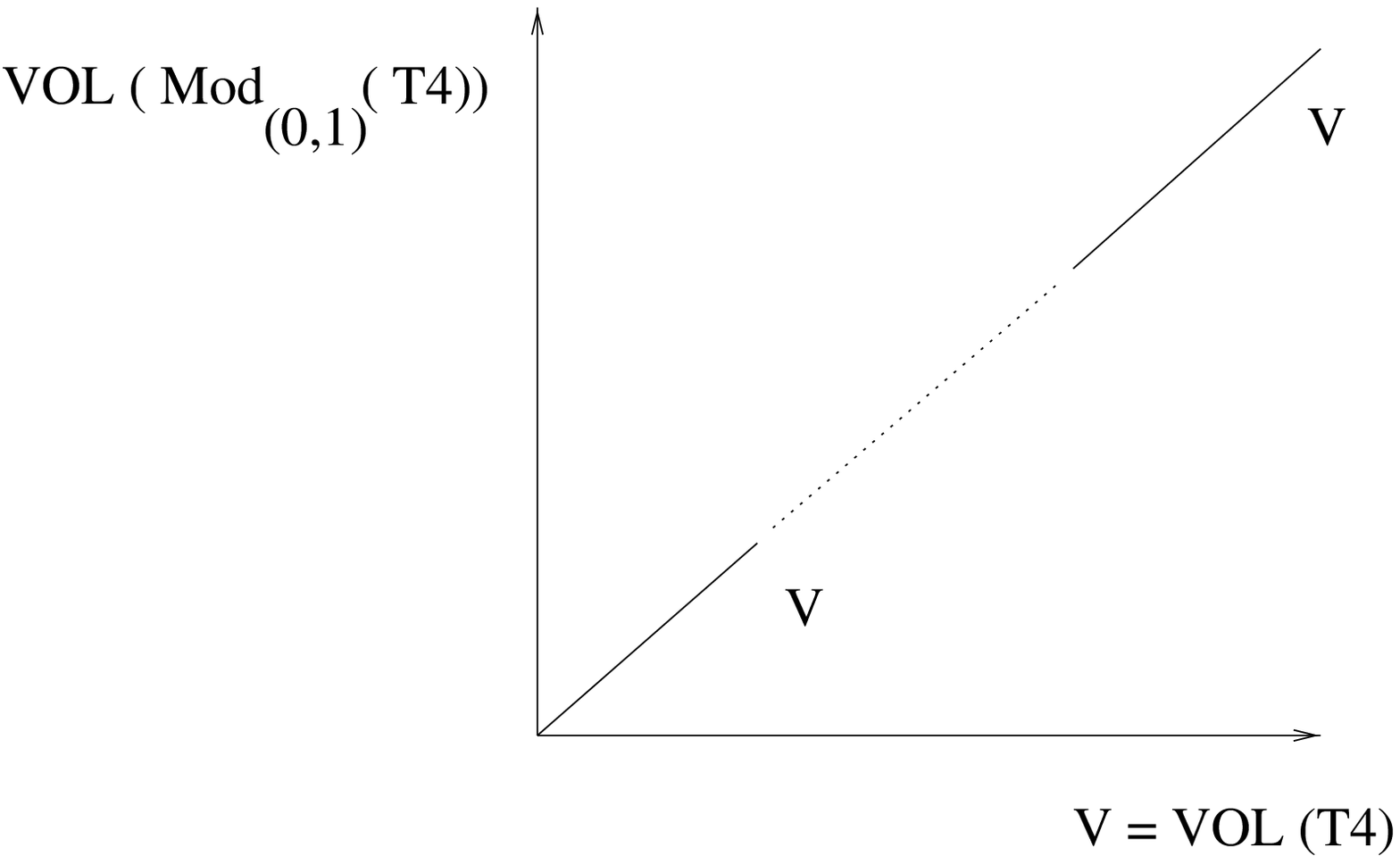} }
 
\medskip

 To summarize: 
\eqn\sum{\eqalign{ 
  &   Vol ( M_{0,1} ( T( V  )  ) )  = V  \hbox{ for large V } \cr  
    &        Vol ( M_{0,1} ( T( V  )  ) )  = V  \hbox{ for small V } \cr }}
 The simplest way to interpolate between 
 these two limits is to take 
 $ Vol ( M_{0,1} ( T ( V ) ) ) = V $ for all $V$. 
 According to this guess
 the volume is equal to $1$ at the self-dual point. This 
 is illustrated in the figure. 
  Also note that the T-duality implies that the K\"ahler parameters 
 obey : 
 \eqn\sumi{  \omega^i  \bigl(  M_{0,1} ( T( V, \omega^i  ) \bigr)
 = \omega^{i} }   
 in both the small and large volume limits.

\subsec{ Zero-brane on $K3$ } 
 Now apply the same considerations to a zero-brane 
 on $K3$ with $B=0$. We certainly have 
 $ Vol ( M_{0,1} ( K ( V ) ) ) = V $ in the large volume limit. 
 Now consider the small volume limit. 
 The T-duality gives us a $K3$ of large volume, 
 and maps the zero-brane to a system of 4-brane 
 and zero-brane. This is because Fourier-Mukai duality  acts simply 
 on the Mukai vector which includes the contribution
 from the curvature of the $K3$. $U(1)$ gauge theory 
 with an instanton describes one 4-brane 
 with no total zero-brane charge, because the charge
 of the $U(1)$ instanton cancels the charge induced from the 
 curvature due to the term $ \int C^{(1)}  \wedge R \wedge R $
 in the $4$-brane action.
 
 Now the moduli space of $U(1)$ gauge theory 
 with a single instanton on $K3$ is just 
 $K3$ with geometry identical to the base space : 
 $Mod_{1,0}(K3) = K3 $. 
 We can see this by the Polchinski D-brane construction 
 of such a system. 
 We can start with the zero-brane and 4-brane being separated
 by a short distance in directions 
 transverse to the $K3$, and take a limit as the zero-brane approaches
 the 4-brane. An open string end-point 
 can end on a zero-brane or a 4-brane. 
 The worldsheet CFT description remains valid, 
 and if $g_s$ is taken to be small, tree level CFT gives
 the correct description.
 From this description a string end-point with Dirichlet boundary conditions 
 along the $K3$ directions can end at any position on the $K3$, 
  which is the location of the zero-brane. An open string connecting 
  zero-brane and 4-brane will have a family of supersymmetric 
 configurations parametrized by the $K3$. 
  The moduli space of such 
  boundary CFTs will clearly the contain the  $K3$ itself.
 There will be boundary marginal operators in this CFT
 which change the location  of the zero-brane, and their two-point 
 functions are determined by the metric on the base space.
 These two-point functions in turn define the 
 metric on the moduli space of the boundary CFTs.
  After we T-dualize 
 a small volume $K3$ of volume $V$ to a $K3$ of large volume $1/V$
 we can use the above argument to show that 
 $Vol ( M_{1,0} ( K( 1/V)  ) )  = 1/V$ in the large 
 volume limit. This implies  that $Vol ( M_{0,1} ( K( V  )  ) )  = 1/V$.

\fig\volmodk{
{Fig.2 } } 
{\epsfxsize3.0in\epsfbox{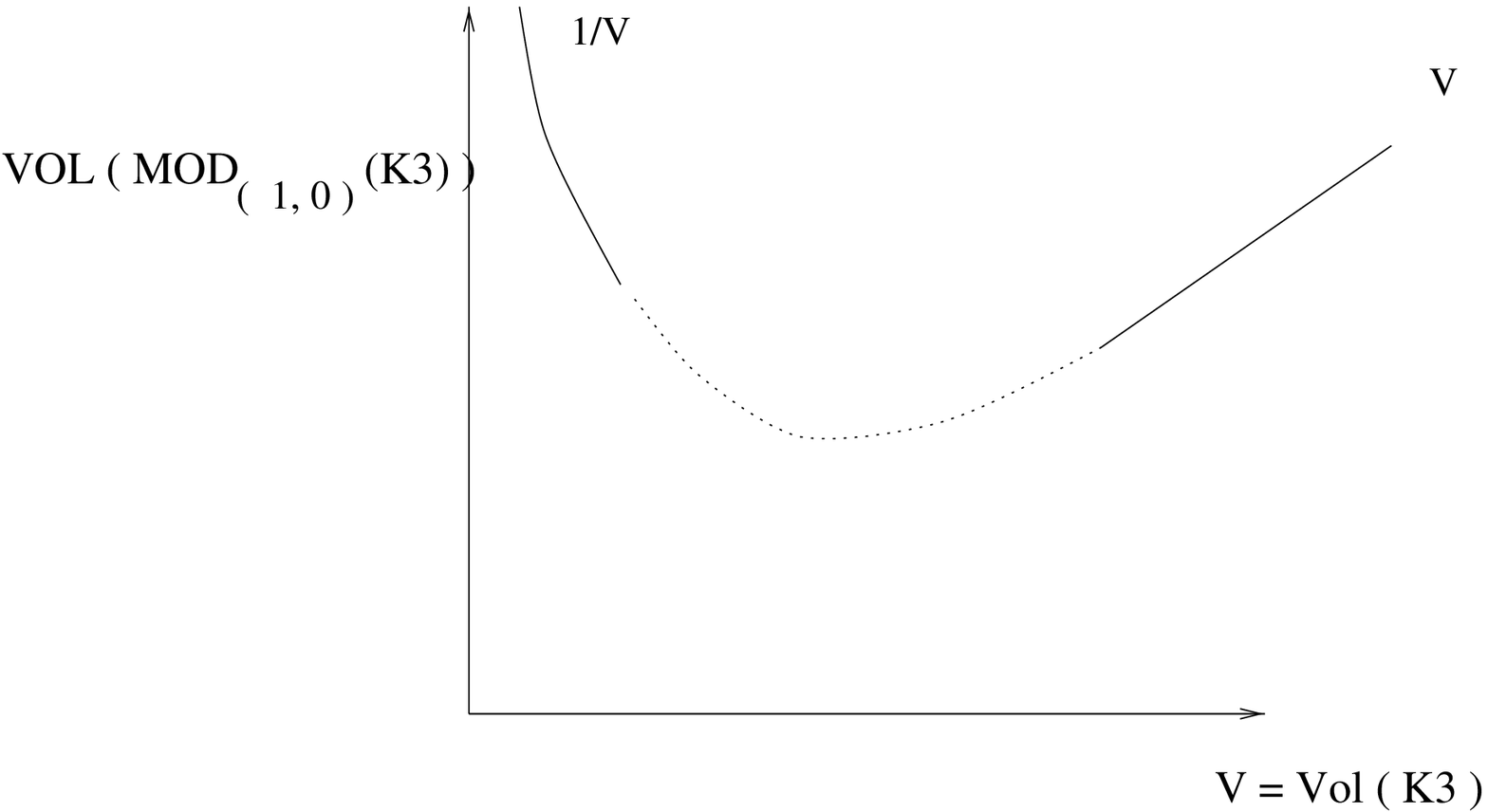} }

\medskip

 To summarize, 
\eqn\sum{\eqalign{ 
  &   Vol ( M_{0,1} ( K( V  )  ) )  = V  \hbox{ for large V } \cr  
    &        Vol ( M_{0,1} ( K( V  )  ) )  = 1/V  \hbox{ for small V } \cr }}
 This is illustrated in the figure. 
 We are further inclined to guess that 
 the volume has one minimum at $V=1 $, where
 $Vol ( M_{0,1} ( K( V  )  ) ) = 1 $. 
 We also deduce from the T-duality
 that the complex structure and K\"ahler 
 parameters of  $M_{0,1} ( K(V) )$ are the same
 in the small volume limit as in the large volume limit.

\subsec{ Extension to $N$ zero branes }

 One obvious extension is to discuss
  $N$ zero branes. 
 We have $S^N(X)$ as the moduli space 
 in the large volume limit for $N$ zero branes. 
 In the limit of small $K3$ we use the duality 
 to map to a system described by $U(N)$ gauge theory 
 with $N$ instantons on a large dual $K3$. The instantons are certainly 
 not ordinary stable sheaves in this case since the 
 Mukai dimension formula gives zero. Presumably it can be proved that
 they are necessarily point-like. So we have a symmetric product 
 of the dual $K3$, which is a $K3$ of large volume. 
 So the moduli space of $N$ zero branes 
  interpolates between being a symmetric 
 product of $N$ $K3$'s of volume $V$ in the large volume 
 limit and being a symmetric product 
 of  $N$ $K3$'s of volume $1/V $ in the small volume limit.

\subsec{ Note on  boundary state definition of the quantum volumes. } 

 We emphasize that the above arguments 
 have used duality to deduce the properties
 of the quantum volume of the moduli space, 
 a quantity which is independently defined 
 in terms of boundary states in the CFTs describing the 
 compactifications.  We are not using the T-duality 
 to define the quantum volumes.

 We can take the coordinates on the end-point
 of the string ending on the zero-brane 
and consider two-point functions
\eqn\twpt{ < x^i(\tau) x^j (\tau^{\prime} > 
= - \alpha^{\prime} g^{ij} log (\tau - \tau^{\prime} )^2 } 
  We can express this correlation function 
 alternatively as : 
\eqn\ders{ 
{  \partial \over \partial \phi^i} { \partial \over \partial \phi^j }  
 \int dg d X e^{ \int S + \int \phi_i \partial_n X^i } } 
 By differentiating we bring down boundary operators. We can express 
 this in terms of boundary states
\eqn\bds{ <0 | B(\phi ) > } 
by taking two derivatives.  Defined in this way we can extend the 
 definition of the metric to abstract CFT's and obtain 
 a definition of the quantum volume. The relevant two-point functions 
 analogous to \twpt\ will have to be computed from more 
 explicit knowledge about the CFT and its boundary marginal
 operators. 
 The T-duality  argument gives a 
prediction for this quantum volume.

\newsec{ Discussion } 
 
 We showed how some elementary facts 
 about simple degenerate instanton moduli spaces
 gives concrete information about quantum volumes. 
 These degenerate instanton moduli spaces 
 appear as sub-strata of larger instanton moduli spaces. 
 A lot of information about such stratifications 
 and their symmetries (acting on the instanton numbers
 and magnetic fluxes characterizing the strata ) 
 are present in BPS mass formulae and associated BPS 
 splittings of the kind studied in detail in \mr\ 
 for the case of $T^4$. The interpretation  of such
symmetries in the case of $K3$ has to take into account the 
 $\alpha^{\prime}$ corrections to instanton moduli spaces
 of the kind discussed here. Information about less degenerate
 smooth strata, 
 e.g of the kind in \yosh\
 could also be used in conjunction with duality 
 to obtain information about quantum corrected geometries.

 The emergence of a
 classical geometry from string-corrected moduli 
 spaces exhibited in Fig. 2. apears somewhat 
 remarkable. 
It is tempting to speculate that 
 some simple rules in string theory 
dictate the appearance of the large volume $K3$. 
For example one might suspect that 
 there is a bound on the volume  seen by any probe
 as one moves in moduli space. Alternatively there might 
 be some constraints on the products of the volumes seen 
 by different probes. To make these speculations more
 precise would require integrating different candidate
 definitions of quantum volumes of cycles, e.g those considered 
 in \grka\ and refs. therein.  Subtleties of the kind discussed
 in \doog\ may have to be dealt with.

\bigskip

 \noindent{\bf Acknowledgements:}
 It is a pleasure to acknowledge 
 discussions with M. Douglas,  J. Harvey, A. Jevicki, S. Kachru,
  A. Lawrence, D. Lowe, M. Mihailescu,  G. Moore, E. Silverstein
 and R. Tatar  on issues related to this note. 
 This work was supported by DOE grant  DE-FG02/19ER40688-(Task A).

\listrefs

\end